\documentclass[onecolumn,showpacs,preprintnumbers,superscriptaddress,amsmath,amssymb,pre]{revtex4}

\usepackage{graphicx}
\usepackage{dcolumn}
\usepackage{bm}

\begin{document}

\preprint{PRE/NewTurbDSI}

\title{New Evidence of Discrete Scale Invariance in the Energy Dissipation of
Three-Dimensional Turbulence: Correlation Approach and Direct
Spectral Detection}

\author{Wei-Xing Zhou}
\affiliation{Institute of Geophysics and Planetary Physics,
University of California, Los Angeles, CA 90095}

\author{Didier Sornette}
\email{sornette@moho.ess.ucla.edu} \affiliation{Institute of
Geophysics and Planetary Physics, University of California, Los
Angeles, CA 90095} \affiliation{Department of Earth and Space
Sciences, University of California, Los Angeles, CA
90095\label{ess} } \affiliation{Laboratoire de Physique de la
Mati\`ere Condens\'ee, CNRS UMR 6622 and Universit\'e de
Nice-Sophia Antipolis, 06108 Nice Cedex 2, France}

\author{Vladilen Pisarenko}
\affiliation{International Institute of Earthquake Prediction
Theory and Mathematical Geophysics, Russian Ac. Sci. Warshavskoye
sh., 79, kor. 2, Moscow 113556, Russia}

\date{\today}

\begin{abstract}

We extend the analysis of [Zhou and Sornette, Physica D 165,
94-125, 2002] showing statistically significant log-periodic
corrections to scaling in the moments of the energy dissipation
rate in experiments at high Reynolds number ($\approx 2500$) of
three-dimensional fully developed turbulence. First, we develop a
simple variant of the canonical averaging method using a rephasing
scheme between different samples based on pairwise correlations
that confirms Zhou and Sornette's previous results. The
second analysis uses a simpler local spectral approach and then
performs averages over many local spectra. This yields stronger
evidence of the existence of underlying log-periodic undulations,
with the detection of more than 20 harmonics of a fundamental
logarithmic frequency $f = 1.434 \pm 0.007$ corresponding to the
preferred scaling ratio $\gamma = 2.008 \pm 0.006$.

\end{abstract}


\maketitle

\section{\label{s:intro} Introduction}

The small-scale physical quantity that has received the most
attention in the context of a phenomenological description of
hydrodynamical turbulence is the dissipation rate of kinetic
energy \cite{Frisch96}. Richardson's picture of turbulent cascades
\cite{Rich22}, in which large eddies break down into smaller ones
receiving a certain fraction of the flux of kinetic energy from
larger scales, is a multiplicative process. The hypothesis that
eddies of any generation are space filling led to the famous K41
theory \cite{K41a,K41b}. A conceptually appealing view, dating
back to Obukhov \cite{O62} and Kolmogorov \cite{K62}, visualizes
the transfer of kinetic energy as a self-similar multiplicative
process. This view is still the point of departure of many
phenomenological intermittency models today. The thought of the
general theory of multifractals was proposed by Mandelbrot
\cite{M74} in the study of certain high-order moments with a
general cascade model of random curling. A phenomenological model
of intermittency, referred to as the $\beta$-model \cite{beta78},
analyzed the multiplicative cascade of the breakup of eddies
without space filling leading to an energy transfer to a smaller
and smaller fraction of the total space. Benzi, Paladin, Parisi \&
Vulpiani \cite{BPPV84} and Frisch \& Parisi \cite{FP85} have
argued for a singular structure and for a multifractal nature of
the energy dissipation in three-dimensional fully-developed
turbulence. Meneveau \& Sreenivasan \cite{MS87} presented a simple
multifractal model, the binomial model for fully developed
turbulence based on the two-scale Cantor set, which is known as
the $p$-model. This model and the specific multifractal facets of
turbulent energy dissipation were investigated experimentally by
\cite{MS91}, using hot wire measurement technique. These works
clarified the phenomenological nature of the multifractality of
turbulent energy dissipation. In particular, the view of the
energy cascade as belonging to the general class of multiplicative
processes opens up the possibility that the scale invariant
properties reported for turbulence flows might be broken in part
into a discrete scale invariance, leading to the observation of
log-periodic corrections to scaling \cite{SorDSI}.

Based on theoretical argument and experimental evidence, Sornette
\cite{Sor98TurbDSI} conjectured that structure functions of
turbulent time series may exhibit log-periodic modulations
decorating their power law dependence. He stressed the need for
novel methods of averaging and proposed a novel ``canonical''
averaging scheme for the analysis of structure factors of
turbulent flows in order to provide convincing experimental
evidence. The strategy was proposed to determine the
sample-specific control parameter $r_c$ and translate it into a
specific ``phase'' in the logarithm of the scale $\ln r$ which,
when used as the origin, allows one to rephase the different
measurements of a structure factor in different realizations.

\begin{figure}
\includegraphics[height=9cm,width=12cm]{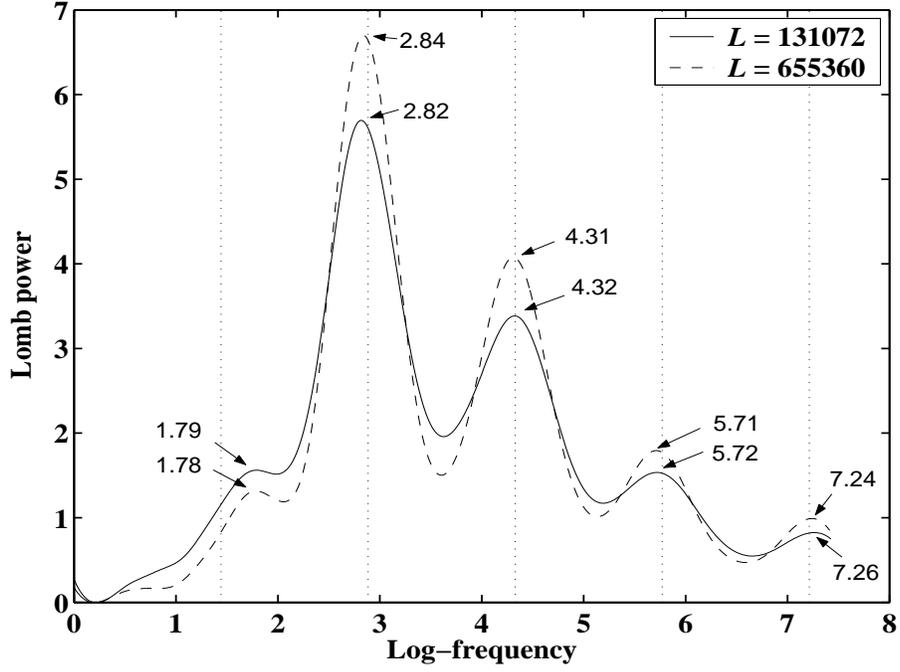}
\caption{\label{Fig1} Average of $2400$ (continuous line) (resp.
$480$ (dashed line)) Lomb periodograms corresponding to $24$
Savitzky-Golay filters per sample. The vertical dashed lines
correspond to log-frequencies equal respectively to $f_k = k/\ln
2$ with $k = 1,\cdots, 5$. These values correspond to increasing
harmonics of a putative fundamental frequency $f_1 = 1/\ln \gamma$
associated with the scaling ratio $\gamma = 2$. (See
Ref.~\cite{TurbDSI} for details.)}
\end{figure}

A first systematic search of discrete scale invariance in
turbulence was carried out in freely decaying two-dimensional
turbulence \cite{Johansen2000b}. The number of vortices, their
radius and separation were found to display log-periodic
oscillations as a function of time with an average log-frequency
of $\sim 4-5$ corresponding to a preferred scaling ratio of $\sim
1.2-1.3$. Most recently, significant evidence for the presence of
intermittent discretely self-similar cascades was obtained from
the analysis of the moments of the energy dissipation rate in
three-dimensional fully developed turbulence \cite{TurbDSI}. The
results of spectral analyses of the canonically averaged local
scaling exponents of high moments of the energy dissipation rate
obtained in \cite{TurbDSI} are summarized in Fig.~\ref{Fig1}. The
fundamental log-frequency is estimated to be $f \approx 1.44$
which suggests a preferred scaling ratio of $\gamma = e^{1/f}
\approx 2$. In \cite{TurbDSI}, a simple multifractal model was
introduced to explain the fact that the log-frequency is
independent of the order of the moments.

As was explained in \cite{TurbDSI}, the difficulty in detecting
log-periodic oscillations (the hallmark of discrete scale
invariance) in various
functions such as moments of the energy dissipation rate stems
from the fact that, while the physics of the cascade may be
universal with an universal scaling factor $\gamma$, the
nucleating scales (translating in phases of reference in the
log-periodicity analysis) have to change from one realization to
another as the cascade may be dynamically triggered from a variety
of scales $\ell_a$ from the integral scale and deep in the
inertial range. Therefore, the corresponding phases are expected
to be non-universal. Averaging a signal over several such
transient cascades will thus wash out the information on
log-periodicity by ``destructive interferences'' of the
log-periodic oscillations, giving them the appearance of noise.

The previous analyses of real 2-D and 3-D turbulence data
addressed this problem by using two versions of the canonical
averaging approach, devised precisely to rephase different
realizations and thus alleviate the problem of the random phases.
Using $n$ realizations of the data, the first version consists in
performing a canonical averaging on the $n$ samples to get an
average realization and then analyzes its spectral properties; the
second version performs a spectral analysis of each of the $n$
samples and then averages the $n$ periodograms. In order to
perform these averages, we need to have different realizations.
For the turbulence data analyzed in \cite{TurbDSI} and revisited
here, a long time series of velocity increments is divided in many
samples, each lasting about $1/5$ of a turn-over time, so that we
can avoid scrambling of the phases by a suitable canonical
averaging as explained below.

The goal of the present note is to
revisit these two schemes and propose two novel and simpler versions. By
simplifying the analysis and by providing independent
implementations, we thus confirm further the presence
of log-periodicity in the moments of the energy dissipation rate, and thus
the existence of discrete scale invariance in 3D fully-developed hydrodynamical
turbulence.

\section{\label{s:Correlation} Rephasing of log-periodicity by
the cross-correlation method}

In the first scheme of canonical averaging, the rephasing operation is
the central part. In \cite{TurbDSI}, the so-called central maxima
approach was applied to the detrended logarithm of the moment
$M_q(\ell/\eta)$ of the dissipation
rate as a function of the logarithm $\ln(\ell/\eta)$ of the scale
$\ell$ (in units of the Kolmogorov microscale  $\eta = 0.195$ mm
\cite{TurbDSI}):
it consists in finding the
maximum of the detrended logarithm of the moment
closest to the middle point of a single sample and set the
logarithm of the corresponding scale (its abscissa) as the origin of
the logarithm of the scales. The new abscissa, called $\Delta$,
becomes sample dependent and is used to superimpose and then average
the detrended moments over the different samples.
This approach was found to
work satisfactorily except that the canonically averaged series present
decreasing amplitudes with increasing $|\Delta|$.

The central
maxima rephasing approach used in \cite{TurbDSI}
determines an absolute sample-dependent
phase which is specific to
each single sample and then rephases all samples.
The alternative idea explored here is based on the remark that
the detection of phases can be performed in relative value, that is,
by estimating the phase shift between all possible pairs of samples.
The cross-correlation function
is used to calculate the phase-shift between two samples \cite{Nikias}.
This relative phase-shift approach thus uses the calculation of
all phase-shifts between pairs of samples to rephase
all samples by choosing arbitrarily one of them as the reference.

As in \cite{TurbDSI}, we use the high-Reynolds
turbulence longitudinal velocity data collected
at the S1 ONERA wind tunnel by the Grenoble group from LEGI
\cite{Anselmet1984}. We use a set of 20 records, each containing $320$
samples in the inertia range,
each sample having $2^{11}$ points. These $2^{11}$ points correspond
approximately to one-fifth of a turn-over time \cite{TurbDSI}.
We calculate the
local moment exponent $\tau \left(q, \ln(\ell/\eta) \right)$ as
the logarithmic derivative of $M_q(\ell/\eta)$ with respect to $\ell/\eta$:
\begin{equation}
\tau \left(q, \ln(\ell/\eta) \right) =
\frac{d\ln[M_q(\ell/\eta)]}{d\ln(\ell/\eta)}~. \label{Eq:tauqell}
\end{equation}
We shall investigate in the sequel $\ell/\eta \in [230, 1843]$
which is well within the inertial range. To obtain $\tau \left(q,
\ln(\ell/\eta) \right)$ from (\ref{Eq:tauqell}),
we used a smoothing low-pass digital
filter, namely, the Savitzky-Golay filter. This filter approximates
the underlying function in a moving window by means of polynomials
of a low degree. The filter coefficients are choosen so that, unlike
other filters, several low-order moments of the filtered function are
preserved,
which is a very desirable property (see \cite{Press1996} for details).
The Savitzky-Golay filters are applied to $\ln[M_q(\ell/\eta)]$
expressed as a function of $\ln(\ell/\eta)$ and the local
derivative is given from the the analytic derivative of the fitted
polynomial \cite{Press1996}. There are two parameters in the
Savitzky-Golay filters, i.e., the width of the sliding window
$N_L+N_R+1$ where $N_L$ and $N_R$ are the numbers of points to the
left and to the right of the investigating point and the order $M$
of fitting polynomials. We used symmetric windows such that $N_L =
N_R$. A total of 24 filters were applied to test the robustness of
the results with respect to the choice of the filters: $N_L = 5,
\cdots, 10$ and $M = 4, \cdots, 7$.

We recall that, in the central maxima approach of \cite{TurbDSI},
one identifies the sample-dependent reduced scale $\ell_c/\eta$
closest to the central point of the scale-interval at which $\tau
\left(q, \ln(\ell/\eta) \right)$ is maximum for each sample. In
contrast, the correlation approach takes an arbitrary sample as
the reference sample and evaluates the phase shift measured by the
cross-correlation function of all other samples relative to this
reference. We then rephase all functions $\tau \left(q,
\ln(\ell/\eta) \right)$ of the variable $\ln(\ell/\eta)$ so that
the cross-correlation function of all sample pairs becomes peaked
at zero lag, after the rephasing has been performed.

\begin{figure}
\includegraphics[height=9cm,width=12cm]{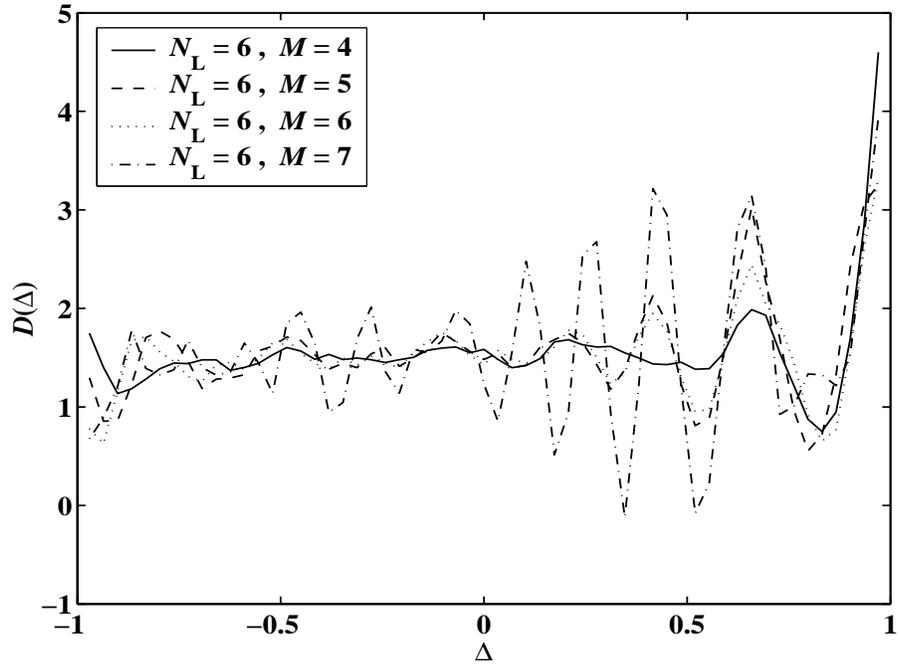}
\caption{\label{Fig2} Four examples of the canonically averaged
variable $D(\Delta)$ as a function of the rephased log-scale
$\Delta$ based on the correlation rephasing approach. The
amplitude of the oscillations increases with increasing $\Delta$.
See text for explanations.}
\end{figure}

We then average these rephased functions $\tau \left(q, \Delta
\right)$ over the $320$ samples. This defines the local
average exponent
\begin{equation}
D(\Delta) = \langle \tau
\left(q, \Delta \right) \rangle~, \label{localderjmsl}
\end{equation}
where the average is performed over the different samples at fixed
$\Delta = \ln(\ell/\eta)- \ln(\ell_c/\eta)$ (where
$\ln(\ell_c/\eta)$ vary from sample to sample and $\ln(\ell/\eta)$
is adjusted accordingly to ensure the re-phasing). Fig.~\ref{Fig2}
shows four examples of $D(\Delta)$ as a function of $\Delta$ based
on the correlation rephasing approach for four different filters
with $N_L = 6$ and $M = 4,5,6,7$ respectively. Notice that the
amplitude of the oscillations decreases with decreasing $\Delta$.
This is in line with the concept \cite{TurbDSI} that, the smaller
is the scale $\ell$, the more complex is the superposition of
intermittent cascades as the number of starting scales for the
cascade proliferate when the nucleating scale departs from the
integral scale. This should lead to a stronger self-averaging
property, as we observe here when $\Delta$ decreases. Furthermore,
since the samples have finite sizes, $D(\Delta)$ is well-defined
only for not too large $|\Delta|$ because there are fewer samples
involved in the averaging for larger $|\Delta|$. In practice, we
will restrict our analysis to $|\Delta|$ smaller than a threshold
$\Delta_m$ taken here approximately equal to $1.0$, as suggested
from a visual inspection of Fig.~\ref{Fig2}.

The next step in the analysis
consists in performing a spectral analysis of $D(\Delta)$. We used a kind of
spectral analysis known under the name of the Lomb spectrum (see
\cite{Press1996} for
details). It is designed to analyze a non-uniformly sampled
functions, fitting them to a sine function of varying frequency on
a non-uniform grid by means of the least square method. This provides
the analog of a standard periodogram. We thus obtain 24 such Lomb periodograms
for each record (of $320$ samples),
one for each filter. We then repeat the above
steps for all the other 19 records of of $320$ samples each and we thus
obtain a total of 480 Lomb periodograms.
Alternatively, we have grouped the 20 records into a single large time
series with 6400 samples on which the
canonical averaging is performed directly. This increased statistics
improves the
significance level of the Lomb peaks in the resultant
periodograms. The comparison of our method on different partitions of the
data allows us to test the stability and robustness of the results.

\begin{figure}
\includegraphics[height=9cm,width=12cm]{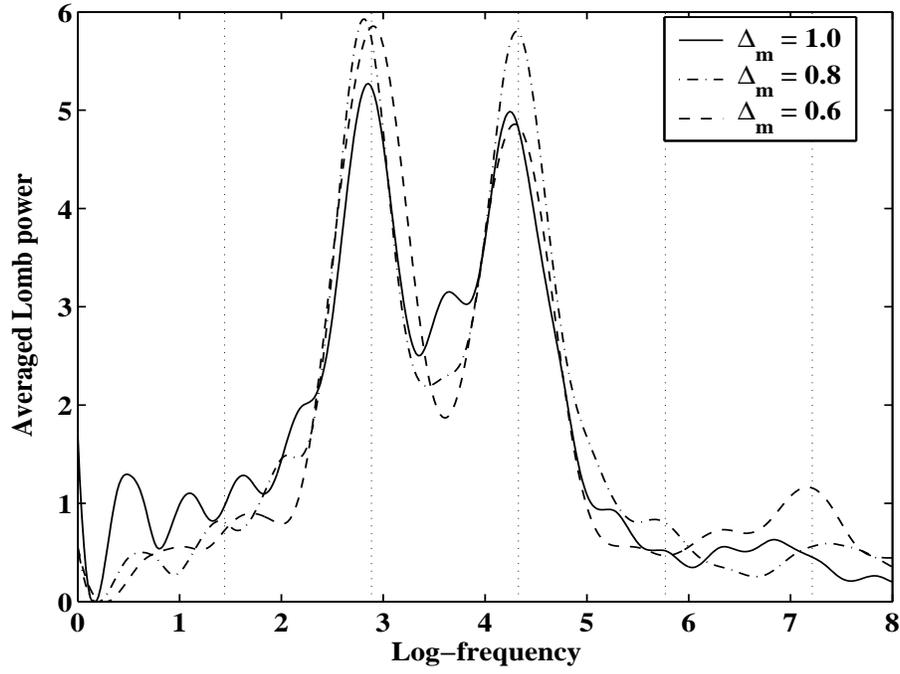}
\caption{Average over all $480$ Lomb periodograms corresponding to
$24$ filters per sample. The vertical dashed lines correspond to
log-frequencies equal to $f_k = k/\ln 2$ with $k = 1,\cdots, 5$.
These values correspond to increasing harmonics of a fundamental
frequency $f_1 = 1/\ln \gamma$ associated with the scaling ratio
close to $\gamma = 2$. } \label{Fig3}
\end{figure}

Having computed the 480 Lomb periodograms (corresponding to 24
Lomb periodograms for each of the 20 records), we perform an
average over all these 480 periodograms. This analysis was carried
out for three different choices of the threshold $\Delta_m = 1.0,
0.8, 0.6$. The averaged Lomb powers are shown in Fig.~\ref{Fig3}.
All the three curves have significant peaks at $f_2 = 2/\ln2$ and
$f_3 = 3/\ln2$, which suggest a fundamental log-frequency $f =
1/\ln2$. The periodogram with $\Delta_m = 1.0$ also possesses many
spurious weaker peaks. The periodogram with $\Delta_m = 0.8$
exhibits the highest Lomb peaks and clearest harmonics. The
difference in peak heights for $\Delta_m = 0.8$ compared with
$\Delta_m = 0.6$ is partially explained by the fact that the
former has more data points. It is indeed known that the Lomb
power is approximately proportional to the number of analyzed
points \cite{LombTests}. The results summarized in Figure
\ref{Fig3} thus provide a strong confirmation of the analysis of
\cite{TurbDSI} on the presence of discrete scale invariance.

\section{\label{s:Direct}Direct detection of log-periodicity by Fast
Fourier Transform}

We now turn to the second method of implementation of the
canonical averaging discussed in the introduction. In a nutshell,
the methods consists in performing directly a spectral analysis of
each sample and then average over the obtained spectra. In order
to improve the signal-to-noise, we first filter the
high-frequencies of the function $\ln[M_q(\ell/\eta)]$ of the
variable $\ln(\ell/\eta)$ by performing an interpolation of the
slightly unevenly sampled third-order moment values
$M_3(\ell/\eta)$ with cubic splines. This provides us with equally
sampled points that are required for the use of a fast Fourier
transform (FFT). We then define the first difference and subtract
the mean which defines the function $\delta \ln [M_3(\ell/\eta)]$
The periodogram of $\delta\ln [M_3(\ell/\eta)$] for each sample is
determined via a fast Fourier transform. We then perform an
averaging over all the periodograms.

\begin{figure}
\includegraphics[height=9cm,width=12cm]{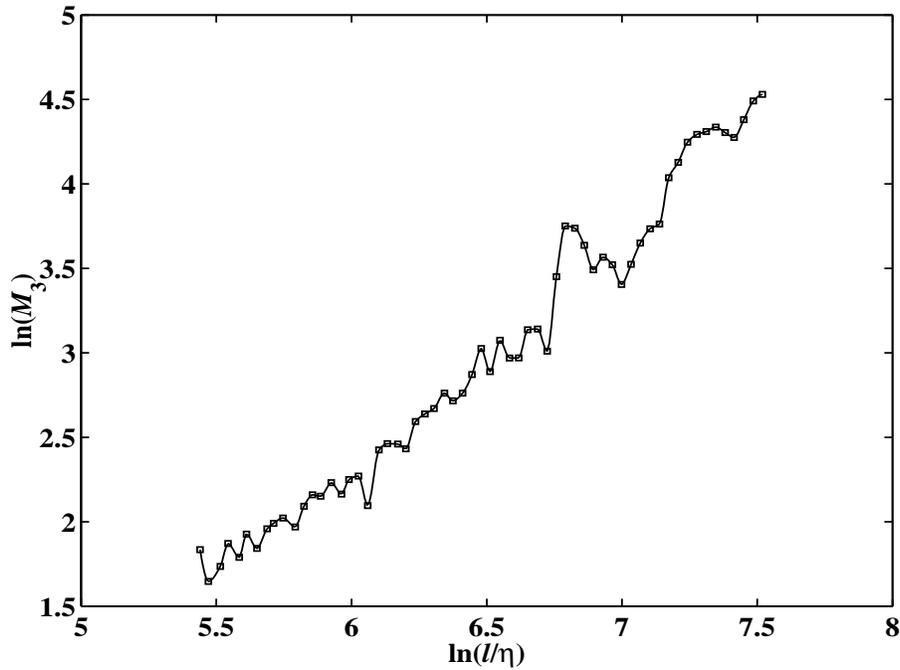}
\caption{\label{Fig4} $\ln(M_3)$ as a function of $\ln(\ell/\eta)$
for a sample (discrete open squares) and its corresponding
interpolation with cubic splines (solid line).}
\end{figure}

\begin{figure}
\includegraphics[height=9cm,width=12cm]{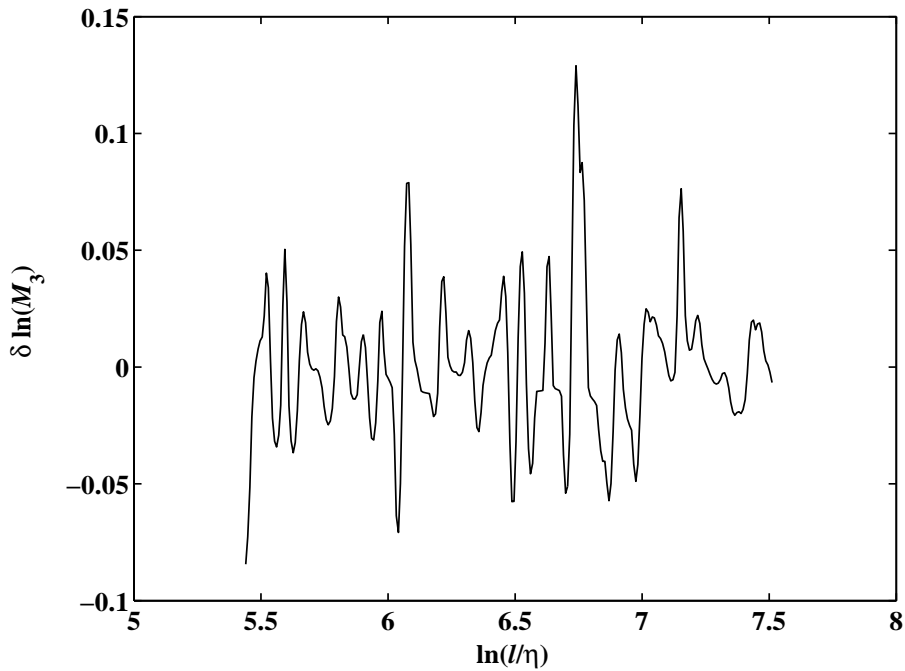}
\caption{\label{Fig5} The zero-mean detrended $\delta\ln(M_3)$ as
a function of $\ln(\ell/\eta)$ for the spline interpolation shown
in Fig.~\ref{Fig4}.}
\end{figure}

\begin{figure}
\includegraphics[height=9cm,width=12cm]{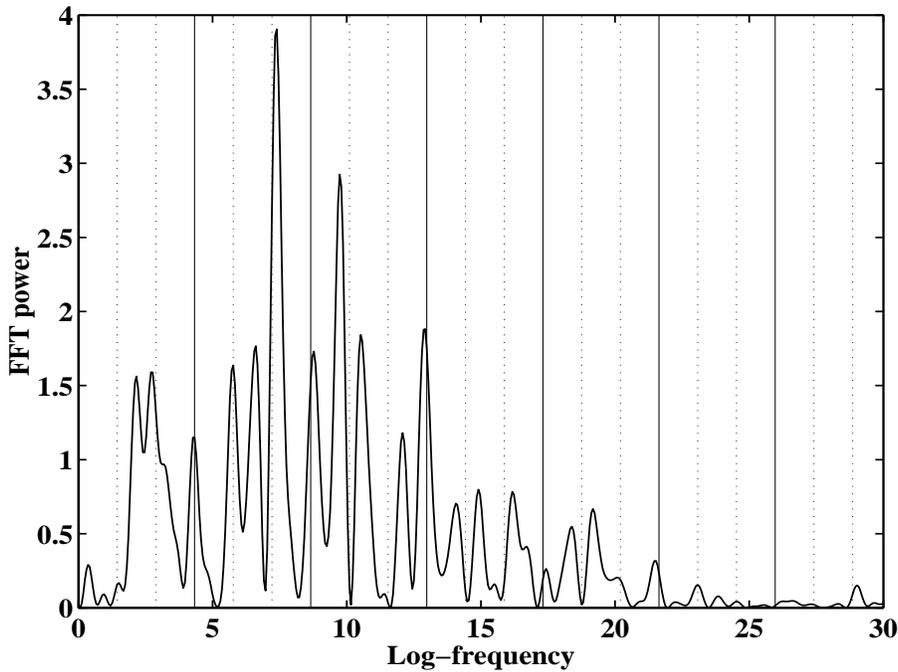}
\caption{\label{Fig6} Fourier periodogram of $\delta\ln(M_3)$
shown in Fig.~\ref{Fig5}. The vertical lines indicate
log-frequencies $k/\ln 2$ with $k = 1,\cdots,21$. There are peaks
close to the vertical lines with $k = 1, 2, 3, 4, 5, 6, 9, 12,
15,...$ and also peaks that can not be integrated as integer
multiples of $1/\ln 2$.}
\end{figure}

Specifically, the calculation of the third-order moment
$M_3(\ell/\eta)$ is done as in Sec.~\ref{s:Correlation} with the
same data sets. We use one record with 320 samples. The number of
the evenly sampled points generated by the spline interpolation is
257, such that we will have 256 points at which the function
$\delta \ln [M_3(\ell/\eta)]$ is sampled. Fig.~\ref{Fig4} shows a
realization of $M_3(\ell)$ for one specific sample. For this,
$\delta_i \ln [M_3(\ell/\eta)]$ is obtained as
\begin{equation}
\delta_i \ln [M_3(\ell/\eta)] = \ln [M_3(\ell_{i+1}/\eta)]-\ln
[M_3(\ell_i/\eta)]-\mu,
\end{equation}
where $i = 1, \cdots, 256$ and $\mu = \langle \ln
[M_3(\ell_{j+1}/\eta)]-\ln [M_3(\ell_j/\eta)] \rangle_j$. The
resulting detrended third-order moment is shown in Fig.~\ref{Fig5}
and its periodogram is depicted in Fig.~\ref{Fig6}. The vertical
lines in Fig.~\ref{Fig6} indicate log-frequencies equal to $k/\ln
2$ with $k = 1,2,\cdots,21$. One can observe well-defined peaks
close to the frequencies corresponding to $k =
1,2,3,4,5,6,9,12,15, 16, 17, 21$. However, there are also peaks in
between these vertical lines. Therefore, the signature of
log-periodicity in Figs.~\ref{Fig4}-\ref{Fig6} is ambiguous.

\begin{figure}
\includegraphics[height=9cm,width=12cm]{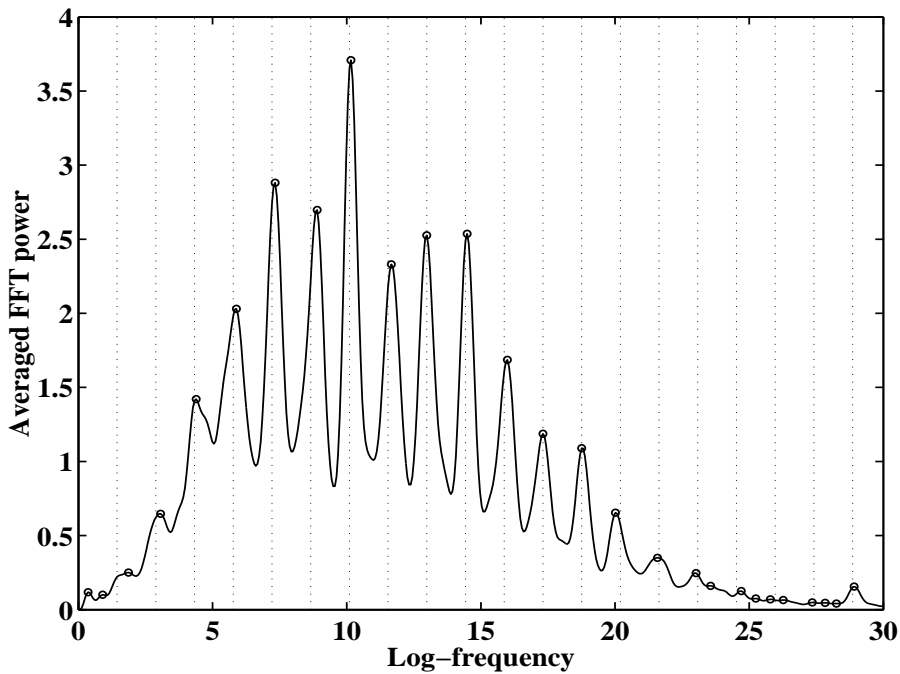}
\caption{\label{Fig7} Average of all Fourier periodograms over 320
samples. The vertical dashed lines correspond to log-frequencies
equal to $f_k = k/\ln 2$ with $k = 1,\cdots, 21$. The peaks are
indicated with open circles. Most of the peaks are very close to
the vertical dotted lines which can be interpreted as harmonics of
$f = 1/\ln2$.}
\end{figure}

We then average over the 320 periodograms, one for each sample,
and get Fig.~\ref{Fig7}. Most peaks at log-frequencies other than
$k/\ln 2$ have been washed out by the averaging procedure. The
vertical dashed lines correspond to log-frequencies $f_k = k/\ln
2$ with $k = 1,\cdots, 21$, which are the harmonics of a
fundamental log-frequency $1/\ln 2$ corresponding to a preferred
scaling ratio equal to $\gamma = 2$. The local peaks defined
preferred frequencies $F_m$ that are shown as open circles. Most
of the peaks ($m = 3, \cdots, 18, 20, 22, 24, 27$) in
Fig.~\ref{Fig7} are very close to the vertical dotted lines which
can be interpreted as harmonics of $f = 1/\ln2$. The background
noise has a bell-shape. Comparing Fig.~\ref{Fig7} with
Fig.~\ref{Fig6}, the averaging of the spectra over the 320 samples
has suppressed peaks that can be interpreted as noise and has
improved remarkably the significance level of the other Lomb
peaks.

\begin{figure}
\includegraphics[height=9cm,width=12cm]{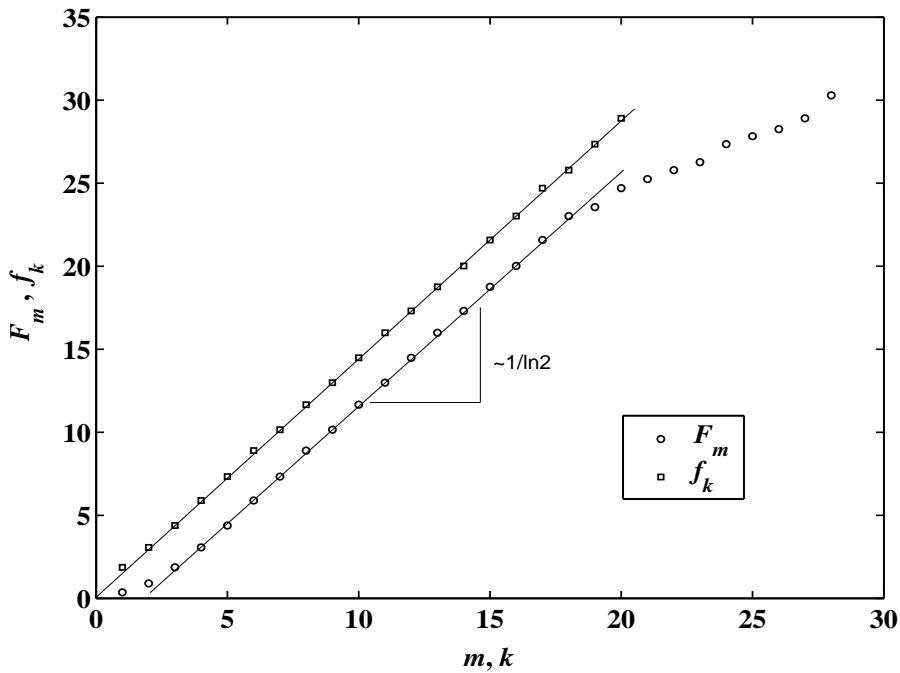}
\caption{\label{Fig8} Determination of the fundamental
log-frequency with a linear fit: the open circles are all peaks
indicated in Fig.~\ref{Fig7} and the open squares are those close
to the vertical lines in Fig.~\ref{Fig7}. Excellent linear fits
are obtained with the slopes $1.420$ ($\circ$) and $1.427$
($\square$), close to the value $1/\ln 2 = 1.4427$.}
\end{figure}

In order to estimate as accurately as possible the numerical value
of the fundamental log-frequency, Fig.~\ref{Fig8} plots with open
circles the log-frequencies $F_m$ of all the peaks found in
Fig.~\ref{Fig7} as a function of the peak sequence number $m$. One
observes a straight line with slope $f = 1.420$ covering 16
harmonics. This value is very close to the theoretic ansatz $f
=1/\ln 2 \approx 1.4427$. An estimation of the uncertainty in the
determination of $f$ is obtained by removing one point in each
data set, and redoing the whole procedure. Then, the standard
deviation calculated over these samples with one removed point
provides a quantification of the error in the determination of
$f$. In order to include more points in the fit, we collected
peaks near the vertical lines in Fig.~\ref{Fig7} with
log-frequencies $f_k$ and plotted $f_k$ with open squares in
Fig.~\ref{Fig8} as well. Note that $f_k$ with $k = 1, \cdots, 20$
corresponds to peaks enumerated $m = 3, \cdots, 18, 20, 22, 24,
27$ mentioned above. Again, a very good linear fit is obtained
with slope $f = 1.427$, which provides a slightly better estimate
of $1/\ln\gamma$. We estimated the slopes of 19 other experimental
records which, together with the slope from the first record, give
the estimate of the fundamental log-frequency $f = 1.434 \pm
0.007$. An alternative estimate of $f$ is the mean of the ratios
$f_k/k$. This gives $f = 1.455 \pm 0.023$. In this estimation, the
first point $(k = 1, f_k = 1.864)$ was discarded.

Overall, we conclude that this new analysis agrees with and strengthens
strongly previous results obtained in Sec.~\ref{s:Correlation} and in
\cite{TurbDSI}
on the observation of log-periodicity. It is also quite
surprising and somewhat gratifying to observe so many harmonics.

\section{Conclusion}
\label{s:concl}

In this paper, we have provided two novel analyses for the search
of log-periodic corrections to scaling in the moments of energy
dissipation of three-dimensional fully developed turbulence. The
first analysis introduces a pairwise cross-correlation approach
for the determination of a phase-shift, which is an alternative of
the central maxima rephasing approach developed in \cite{TurbDSI}
and is more natural and straightforward to implement. The
resulting averaged Lomb periodogram have two significant peaks
near log-frequencies equal to $2/\ln2$ and $3/\ln2$ and low peaks
near other $k/\ln2$ with $k = 1, 4, 5$. We have checked that these
results are robust when varying the range of scales over which the
analysis is performed.

The second method uses a more straightforward sample-by-sample
spectral analysis, with a simple cubic spline interpolation
(rather than the more delicate Savitzky-Golay filters) and a
standard FFT. Averaging over the hundreds of spectra provides a
strong log-periodic signals with more than 20 harmonics of what
appears to be a fundamental logarithmic frequency $f = 1.434 \pm
0.007$. The corresponding preferred scaling ratio is thus
estimated to be $\gamma = 2.008 \pm 0.006$. This novel analysis
lifts possible remaining worries that one could keep that the
results of \cite{TurbDSI} could be somehow due to the use of the
Savitzky-Golay filters.

The present work further confirms the presence of log-periodicity
in the energy transfer of fully developed turbulence, as well as
provide additional evidence of the presence of multifractality in
the energy dissipation \cite{MS91}. However, the preferred scaling
ratio $\gamma \approx 2$ found here does not necessarily imply
that the correct model for turbulence is the binomial multifractal
cascade \cite{Lapidus}. We ascribe our results as an imprint of
intermittent cascades from discrete hierarchical dissipation in
turbulence  \cite{TurbDSI}.

\begin{acknowledgments}
The experimental turbulence data obtained at ONERA Modane were
kindly provided by Y. Gagne. We are grateful to J. Delour and
J.-F. Muzy for help in pre-processing these data and to A.
Helmstetter for pointing out the correlation approach. This work
was partially supported by NSF-DMR99-71475 and the James S. Mc
Donnell Foundation 21st century scientist award/studying complex
system.
\end{acknowledgments}

\pagebreak \clearpage

\end{document}